\documentclass[amsmath,amssymb,aps,twocolumn]{revtex4-1}

\usepackage{graphicx}% Include figure files
\usepackage{dcolumn}% Align table columns on decimal point
\usepackage{bm}% bold math
\usepackage{hyperref}
\hypersetup{colorlinks=true,linktoc=all,citecolor=blue,linkcolor=blue}

\def\beq{\begin{equation}}
\def\eeq{\end{equation}}
\def\ba{\begin{align}}
\def\ea{\end{align}}

\newcommand{\tcite}[1]{~\cite{#1}}
\newcommand{\braket}[1]{\langle #1 \rangle}
\newcommand{\braketd}[1]{\langle\!\langle #1 \rangle\!\rangle}

\begin{document}

\title{Mixed-parity superconductivity near Lifshitz transitions in strongly spin-orbit-coupled metals}% Force line breaks with \\

\author{Matthew J. Trott}
\affiliation{SUPA, School of Physics and Astronomy, University of St Andrews, North Haugh, St Andrews, Fife KY16 9SS, United Kingdom}
\author{Chris A. Hooley}
\affiliation{SUPA, School of Physics and Astronomy, University of St Andrews, North Haugh, St Andrews, Fife KY16 9SS, United Kingdom}

\date{27th November 2019}

\begin{abstract}
\noindent
We consider a strongly spin-orbit-coupled metal, one of whose Fermi surfaces is close to a Lifshitz (topological) transition.  Via a renormalization group analysis of the square-lattice Hubbard model with strong Rashba spin-orbit coupling, we show that such a metal is generically unstable to the formation of mixed-parity superconductivity with a helical triplet component.
\end{abstract}

\maketitle

\noindent
\section{Introduction} 
Topological superconductivity is at the forefront of modern investigations in materials physics due in part to its potential for realizing topological quantum computation via localized Majorana zero modes\tcite{sarma2015}. In order to obtain non-trivial topology the superconductivity must be of an unconventional form, with spin-triplet Cooper pairs carrying non-zero angular momentum\tcite{sato2017}.  Such unconventional superconductivity is thought to arise from spin-fluctuation-mediated pairing, distinct from the phonon-mediated mechanism which is found in the majority of superconducting materials\tcite{kohn1965}. 

The search for materials that have the required topological characteristics is ongoing. A class of materials of great interest are those with non-centrosymmetric or non-symmorphic crystal structures. The broken inversion symmetry of the non-centrosymmetric materials allows for antisymmetric spin-orbit coupling\tcite{smidman2017}.  This leads to a mixing of spin-singlet and spin-triplet Cooper pairs\tcite{gorkov2001}. The non-centrosymmetric material CePt$_3$Si is a candidate for an ($s+p$)-wave superconducting state\tcite{bauer2004,samokhin2004}. It is also believed that Li$_2$Pt$_3$B exhibits spin-triplet superconductivity\tcite{nishiyama2007}. 

 Other spin-triplet candidates include cuprate thin films grown on a substrate. These films have an induced Rashba spin-orbit interaction which leads to a ($d+p$)-wave pairing state\tcite{yoshida2016,takasan2017}. Unconventional superconductivity is also thought to arise at oxide interfaces\tcite{scheurer2015}. Additionally, there are proposals to engineer unconventional superconductors via superlattices of organic molecules on superconducting substrates\tcite{lu2016}. 

The unifying physics within the proposed mixed-parity superconductivity materials is the presence of antisymmetric spin-orbit coupling. This coupling induces spin flips during scattering processes, altering how the superconductivity is formed, and leads to an enhancement of the triplet component of the superconducting order parameter\tcite{yokoyama2007}.

A simple model that captures both the spin-orbit interaction and the electron-electron repulsion needed for unconventional pairing is the square-lattice Hubbard model with an additional Rashba spin-orbit coupling term. The Rashba term splits the underlying tight-binding band into two bands with non-trivial spin textures. This model, known as the Rashba-Hubbard model, has been investigated in several previous studies. Early analysis of the extended Rashba-Hubbard model using the random phase approximation (RPA) suggested mided-parity superconductivity\tcite{yokoyama2007,shigeta2013}. Later RPA studies found regions of $d$-wave and $f$-wave superconductivity dominating\tcite{greco2018,ghadimi2019}.  Dynamical mean-field-theory studies found a mixed-parity state of ($d+p$)-wave superconductivity\tcite{lu2018}.
 
An important feature of the two-dimensional Hubbard model is the presence of saddle points in the tight-binding dispersion. The saddle points lead to a van Hove singularity that occurs when the system is doped through a Lifshitz transition, a Fermi surface transition between an open and closed Fermi surface, or the connection of two or more Fermi surface pockets\tcite{lifshitz1960,volovik2017}. At a van Hove singularity the density of states diverges logarithmically, with electrons around saddle points in the dispersion giving rise to the divergence. A patch approximation can then be constructed around the van Hove saddle points to examine the possible Fermi surface instabilities of the system\tcite{schulz1987,lederer1987,furukawa1998}.
 
In the Rashba-Hubbard model there are two filling fractions at which the Fermi surface touches the van Hove points. At each such filling all the van Hove points lie on one Fermi surface sheet such that the predominant contribution to the susceptibilities comes from a single helicity band in the low-energy limit.

When electron-electron interactions are stronger than or comparable to the Rashba coupling, significant scattering between the two helicity bands is expected.  However, when the electron-electron interactions are much weaker than the Rashba coupling, low-energy scattering is predominantly intra-band, and that is the limit we shall consider here.

In this Article, following\tcite{yao2015,nandkishore2012,huang2016}, we construct a patch renormalization group (RG) scheme applicable near the two van Hove fillings of the Rashba-Hubbard model. In the limit of weak coupling we perform a one-loop RG to find the leading instability of the system. We show that mixed-parity superconductivity arises naturally from this scheme, demonstrating the importance of the spin helicity structure when considering spin-orbit coupled materials.

\section{Model and methods}
We consider a square lattice tight-binding model,
\begin{align}\label{rhham}
H_0=&-t\sum_{\braket{i,j},s}c^{\dagger}_{is}c_{js}-t'\sum_{\braketd{i,j},s}c^{\dagger}_{is}c_{js}-\mu\sum_{i,s} c^{\dagger}_{is}c_{is}
\nonumber\\&+i\upsilon\sum_{\braket{i,j},s,s'} \left[(\bm{\sigma}\times\mathbf{a}_{ij})\cdot\mathbf{\hat{z}}\right]_{ss'} c^{\dagger}_{is}c_{js'}.
\end{align}
$\braket{i,j}$ and $\braketd{i,j}$ denote nearest-neighbor and next-nearest-neighbor hopping with hopping strengths $t$ and $t'$ respectively. The spin orientations are denoted $s,s' \in \left\{ \uparrow,\downarrow \right\}$. $\upsilon$ is the Rashba spin-orbit coupling strength,  $\bm{\sigma}=(\sigma_x,\sigma_y,\sigma_z)^T$ the vector of Pauli matrices, and $\mathbf{a}_{ij}$ denotes the unit vectors between nearest-neighbor sites.

\begin{figure}[t]
	\centering
	\includegraphics[width=0.49\columnwidth]{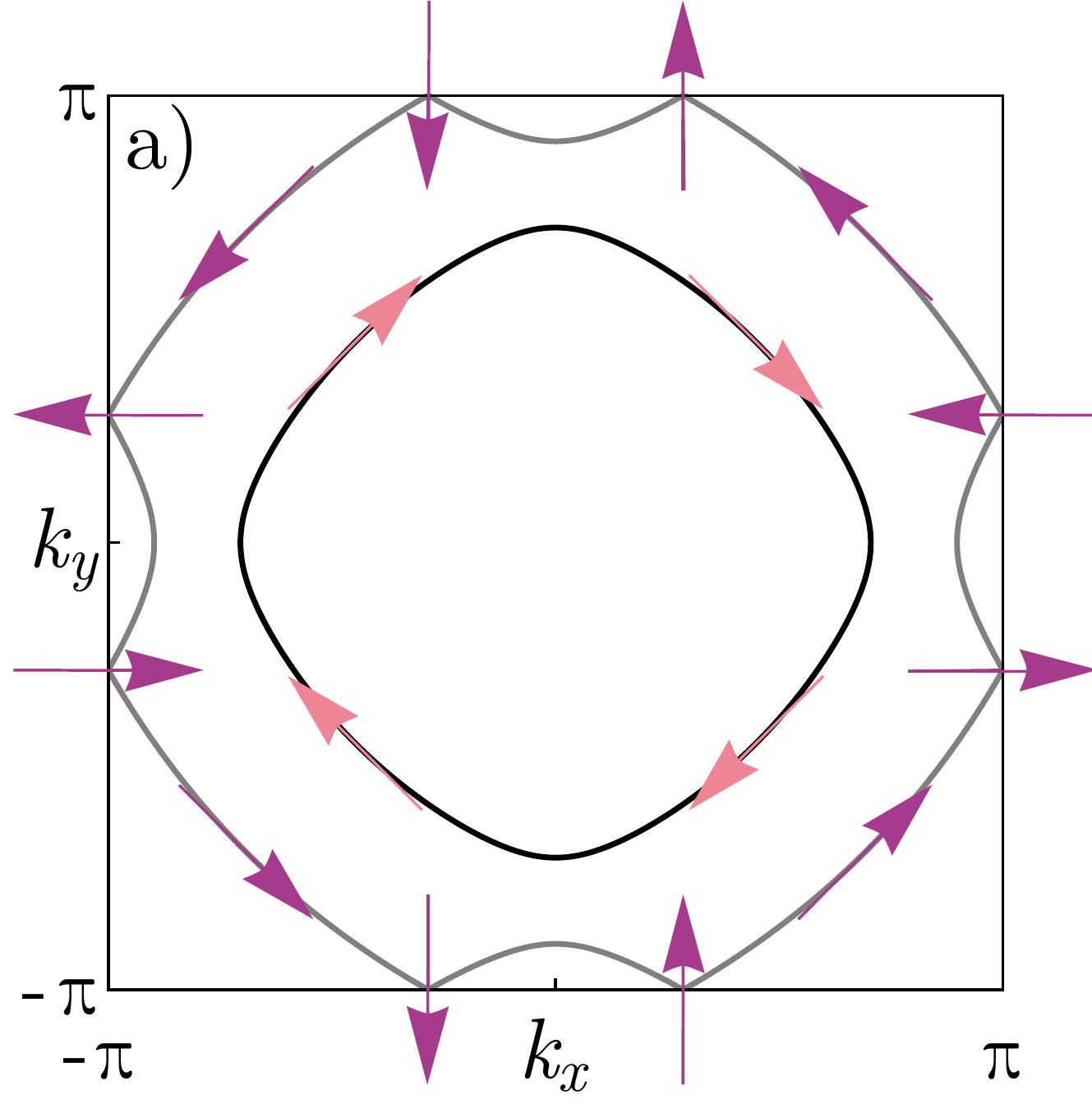}
	\includegraphics[width=0.49\columnwidth]{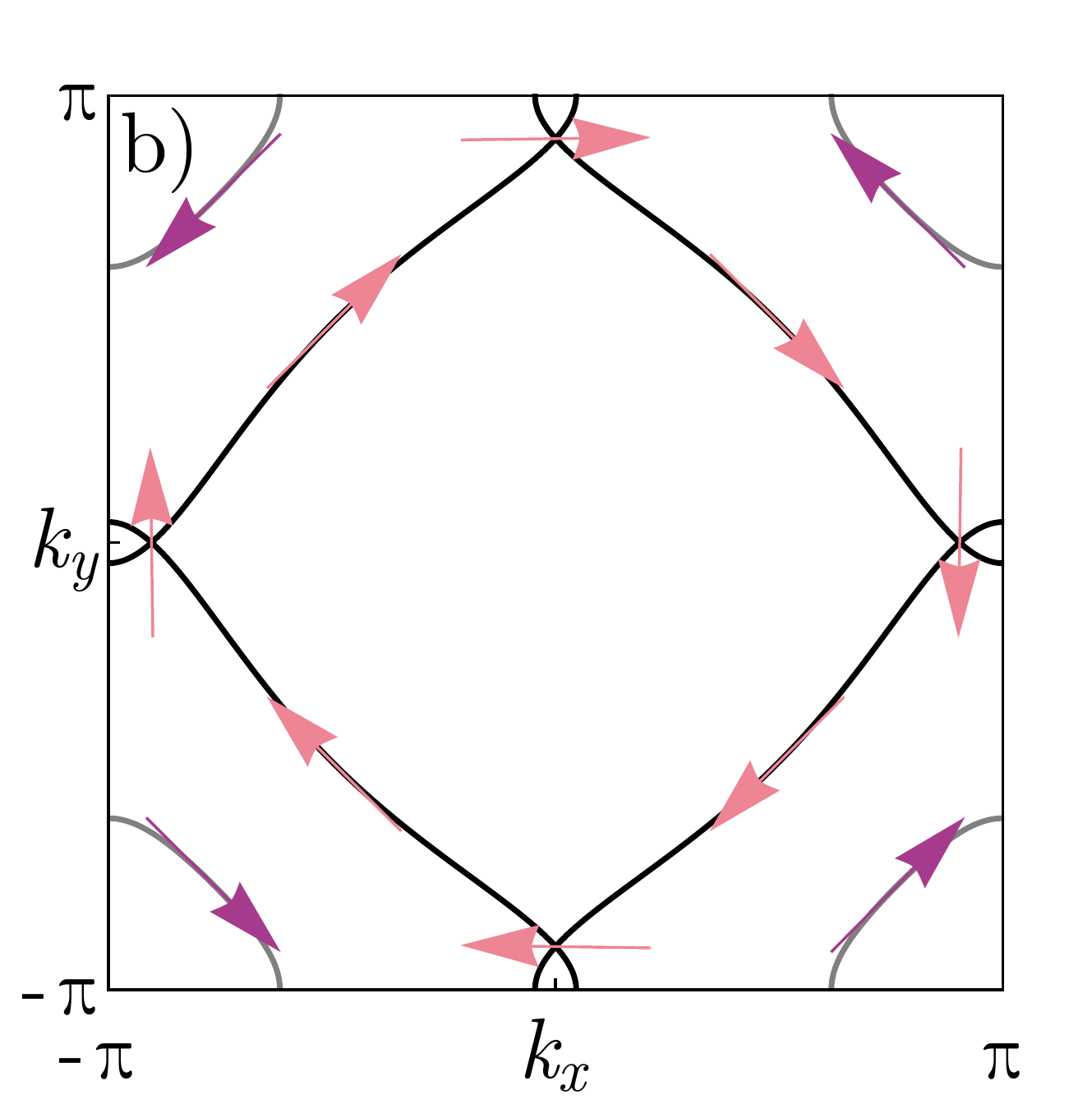}
	\caption{a) The two Fermi-surface sheets of the non-interacting Rashba-Hubbard model when the chemical potential is $\mu_{-}$, the value at which van Hove singularities occur on the outer sheet.  b) The same, but for the chemical potential $\mu_{+}$, the value at which van Hove singularities occur on the inner sheet.  In both panels arrows indicate the spin direction of the helicity eigenstate on that sheet of the Fermi surface.  The parameters used are $t=1$, $t'=0.3$, and $\upsilon=0.5$.}
	\label{helicity}
\end{figure}

The Hubbard interaction term is
\beq 
V_{\text{int}}=\frac{U}{2}\sum_{s,s'}\sum_{k_1k_2k_3k_4}\delta_{k_1+k_2-k_3-k_4}c^\dagger_{k_1s}c^\dagger_{k_2s'}c_{k_3s'}c_{k_4s},
\eeq
describing a contact interaction which is repulsive for $U>0$ and attractive for $U<0$. The interacting Hamiltonian $H$ is given by $H_0+V_{\text{int}}$. 
 
 Spin-orbit coupling breaks the spin degeneracy of the non-interacting bands and splits them into two with opposite helicities. After a unitary transformation to the helicity basis the non-interacting Hamiltonian (\ref{rhham}) becomes  $H_0=\sum_{k,\alpha}\xi^\alpha_kc^\dagger_{k\alpha}c_{k\alpha}$, with the two helicities denoted by Greek indices $\alpha\in\{+,-\}$.
\beq 
\xi^\pm_k=\epsilon_k-\mu\pm2\upsilon\sqrt{\sin^2k_x+\sin^2k_y}
\eeq
with $\epsilon_k$ the next-nearest-neighbor Hubbard model dispersion
\beq 
\epsilon_k=-2t(\cos k_x+\cos k_y)-4t'\cos k_x\cos k_y.
\eeq
Here and henceforth we set the lattice spacing to unity. The eigenvectors for the helicity bands are
\beq 
|\eta_\pm(\mathbf{k})\rangle=\frac{1}{\sqrt{2}}\begin{pmatrix}1\\ \pm e^{i\theta(\mathbf{k})}\end{pmatrix}
\eeq
with
\beq 
 e^{i\theta(\mathbf{k})}=\frac{\sin k_y-i\sin k_x}{\sqrt{\sin^2k_x+\sin^2k_y}}.
\eeq
The spin orientations along the helicity bands are shown in Fig. \ref{helicity}. The operator mapping to the helicity basis is given by
\beq 
c_{k\pm}=\frac{1}{\sqrt{2}}(c_{k\uparrow}\pm e^{-i\theta(\mathbf{k})}c_{k\downarrow}).
\eeq

The bare anti-symmetrized Hubbard interaction term $V_{\text{int}}$ in the helicity basis becomes\tcite{wang2014}
\begin{eqnarray} \label{antiU}
V_{\text{int}} & = & \frac{U}{16}\sum_{\alpha\beta\gamma\delta}\sum_{k_1k_2k_3k_4}\delta_{k_1+k_2-k_3-k_4}( \alpha e^{-i\theta(\mathbf{k}_1)}- \beta e^{-i\theta(\mathbf{k}_2)}) \nonumber \\ & & \qquad \times (\delta e^{i\theta(\mathbf{k}_4)}- \gamma e^{i\theta(\mathbf{k}_3)}) c^\dagger_{k_1\alpha}c^\dagger_{k_2\beta}c_{k_3\gamma}c_{k_4\delta}.
\end{eqnarray}

The two van Hove fillings are located at chemical potentials $\mu_+$ and $\mu_-$, given by 
\begin{multline}
\mu_\pm  =  \pm2\left(-t+\frac{(t\pm2t')^2}{\sqrt{(t\pm2t')^2+\upsilon^2}}+\right.\\ \left.+\upsilon\sqrt{1-\frac{(t\pm2t')^2}{(t\pm2t')^2+\upsilon^2}}\right).
\end{multline}
 At each van Hove filling there are four van Hove points in the Brillouin zone, as shown in Fig. (\ref{helicity}). The filling $\mu_+$ is reminiscent of the scenario proposed by Yao and Yang\tcite{yao2015}, denoted a type-II van Hove singularity, with saddle points located away from the Brillouin zone edge. In the $\mu_-$ case the type-I van Hove point\tcite{schulz1987,lederer1987,furukawa1998} splits into two along the Brillouin zone edge. We call this the edge van Hove scenario. The van Hove points lie at $\mathbf{K}_{1,2}=(\mp \Pi^+,0)$, $\mathbf{K}_{3,4}=(0,\mp \Pi^+)$ for filling $\mu_+$ and $\mathbf{K}_{1,2}=(\mp \Pi^-,\pi)$, $\mathbf{K}_{3,4}=(-\pi,\mp \Pi^-)$ for filling $\mu_-$, where
 \beq 
 \Pi^\pm=\arccos\left(\mp\frac{t\pm2t'}{\sqrt{(t\pm2t')^2+\upsilon^2}}\right).
 \eeq

\begin{figure}[t]
	\centering
	\begin{tabular}{c|c|c}
	\includegraphics[width=0.32\columnwidth]{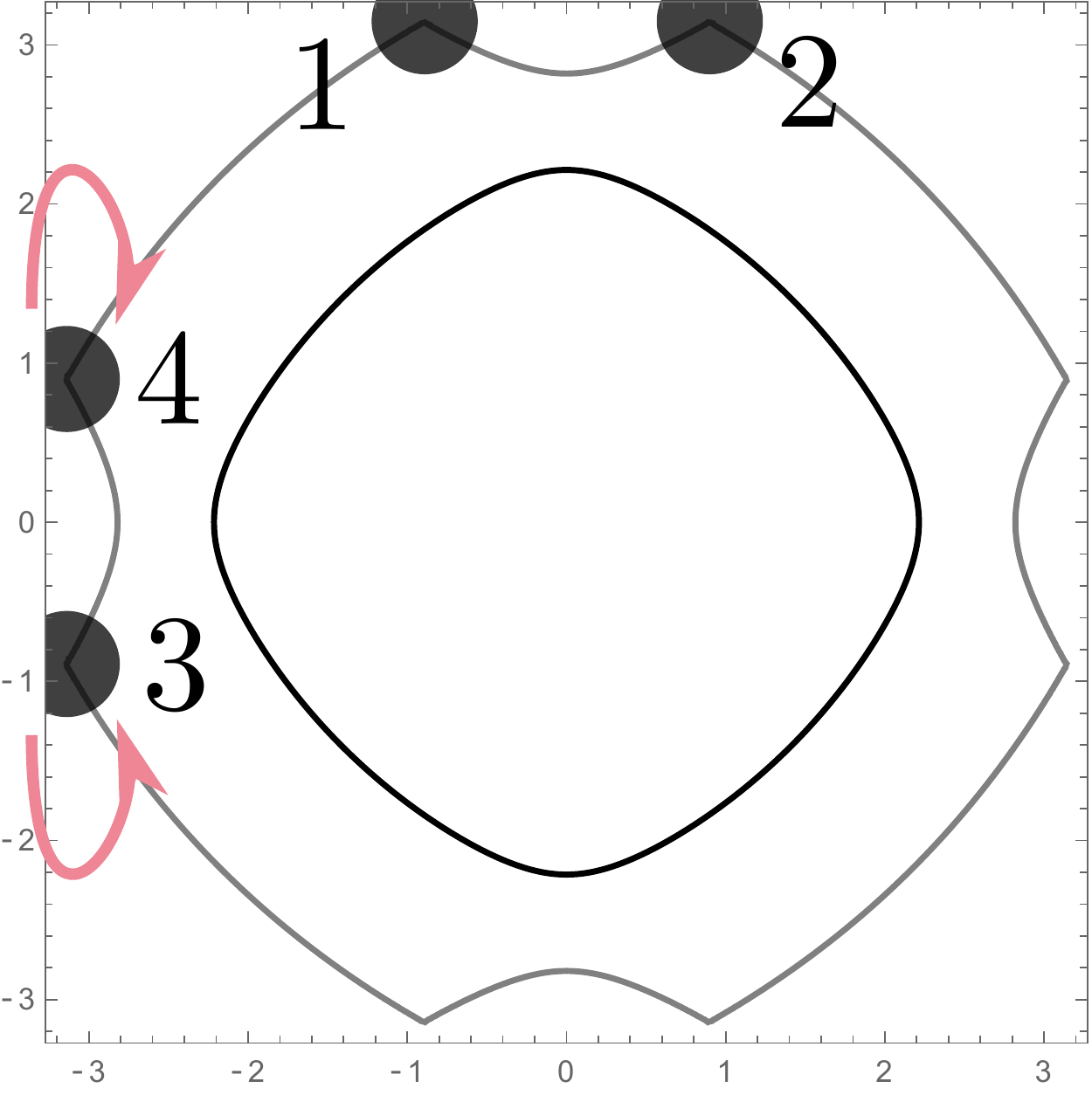}&	
	\includegraphics[width=0.32\columnwidth]{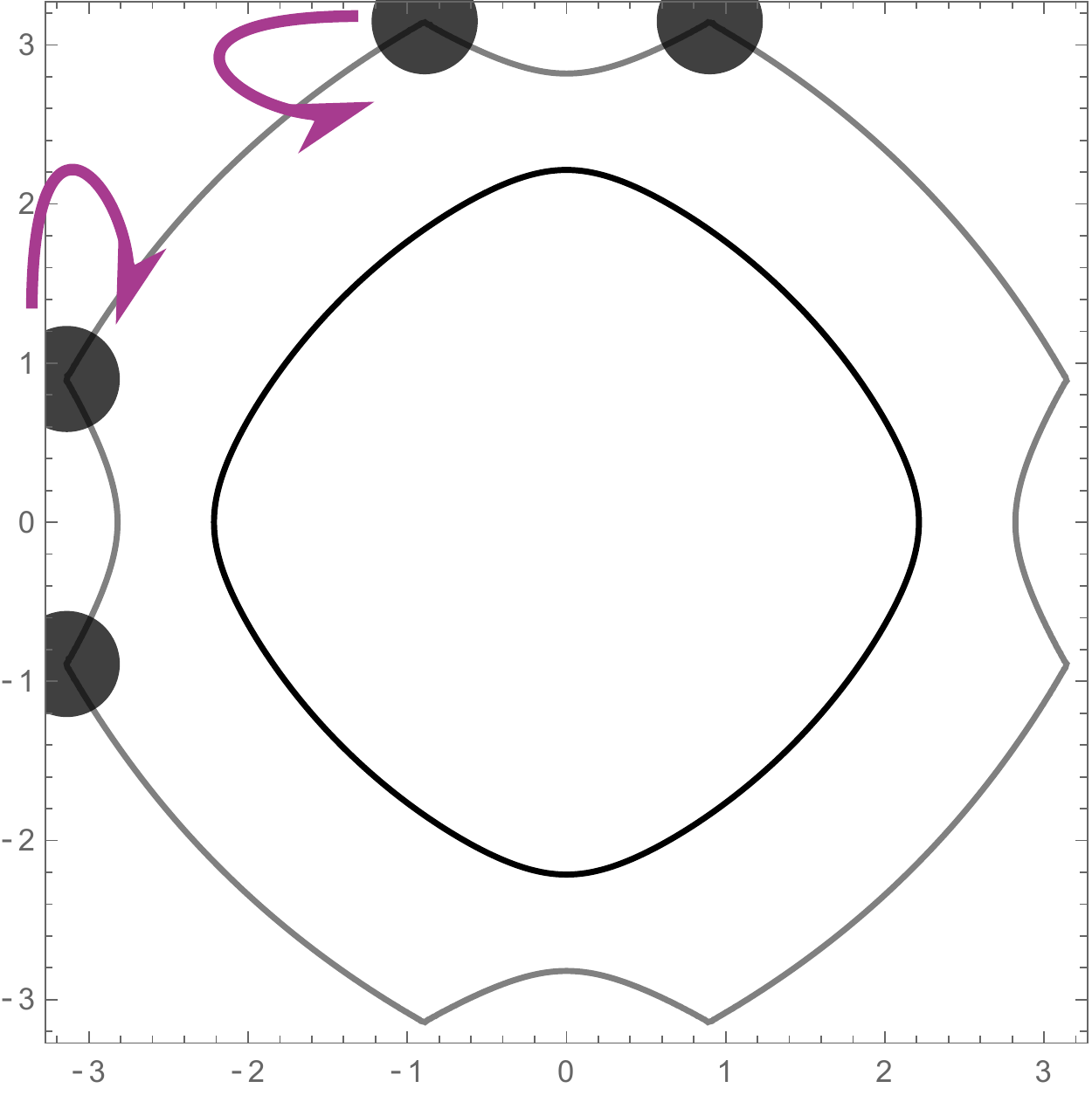}&	
	\includegraphics[width=0.32\columnwidth]{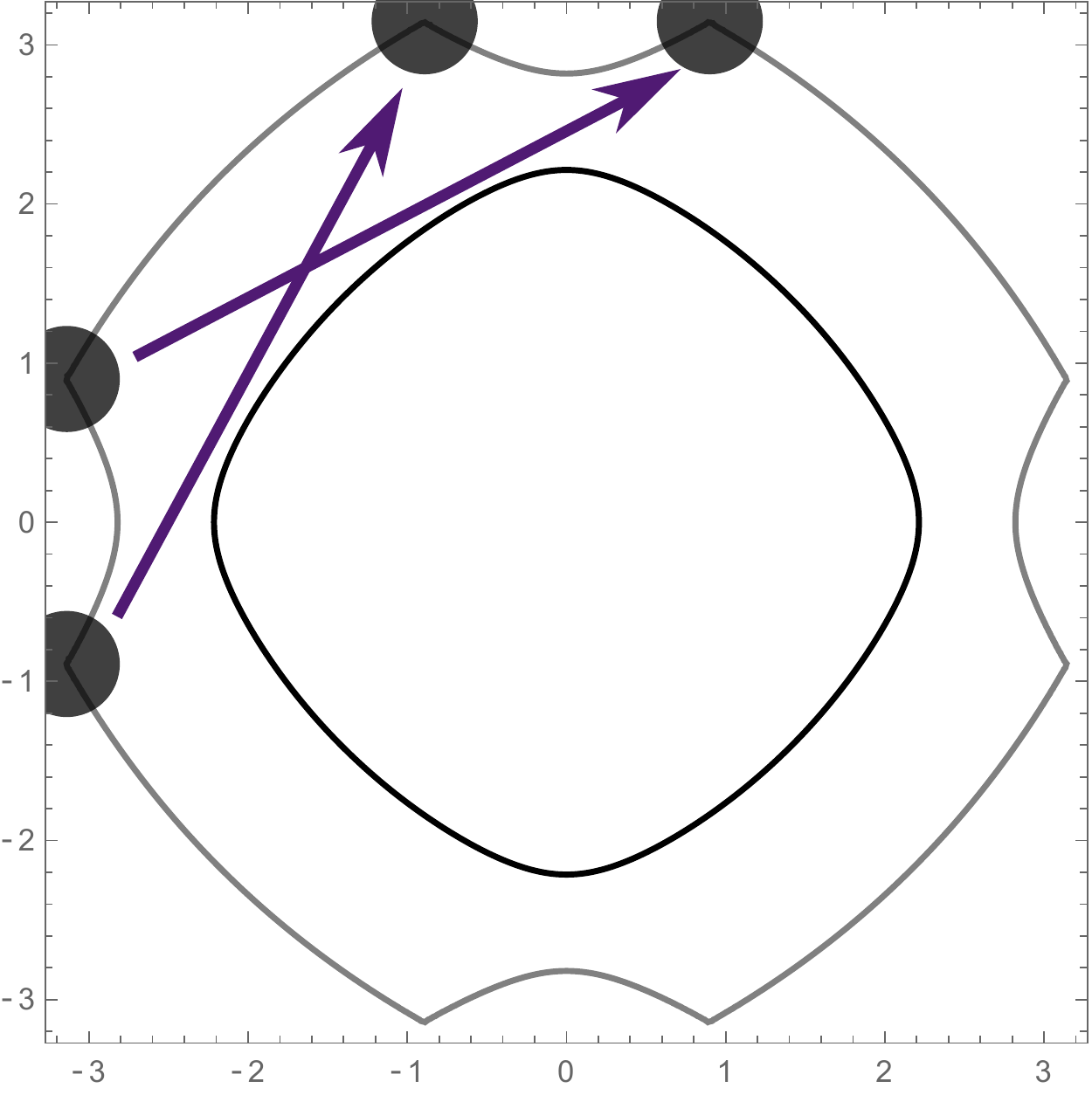}\\
	\includegraphics[width=0.32\columnwidth]{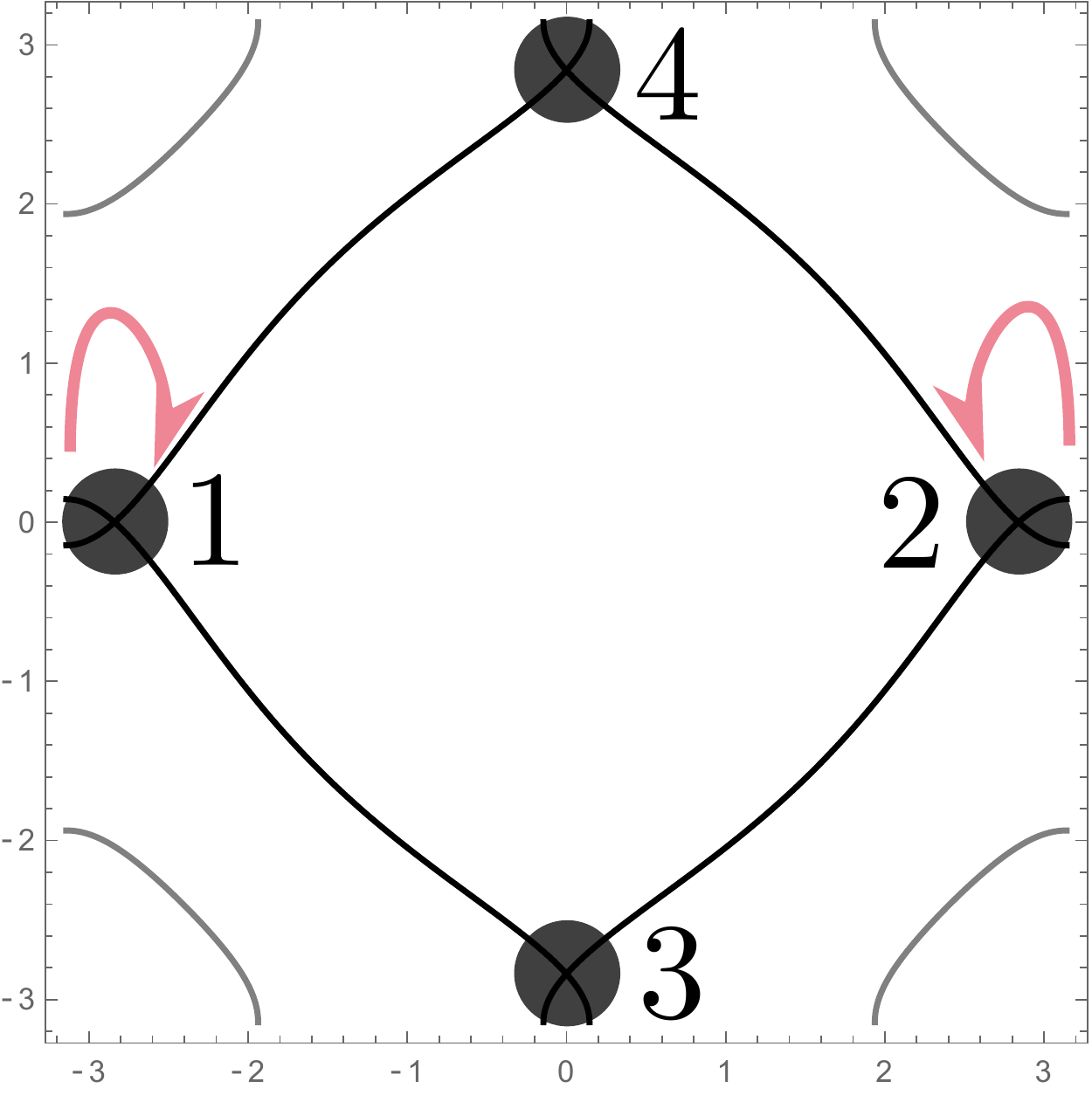}&	
	\includegraphics[width=0.32\columnwidth]{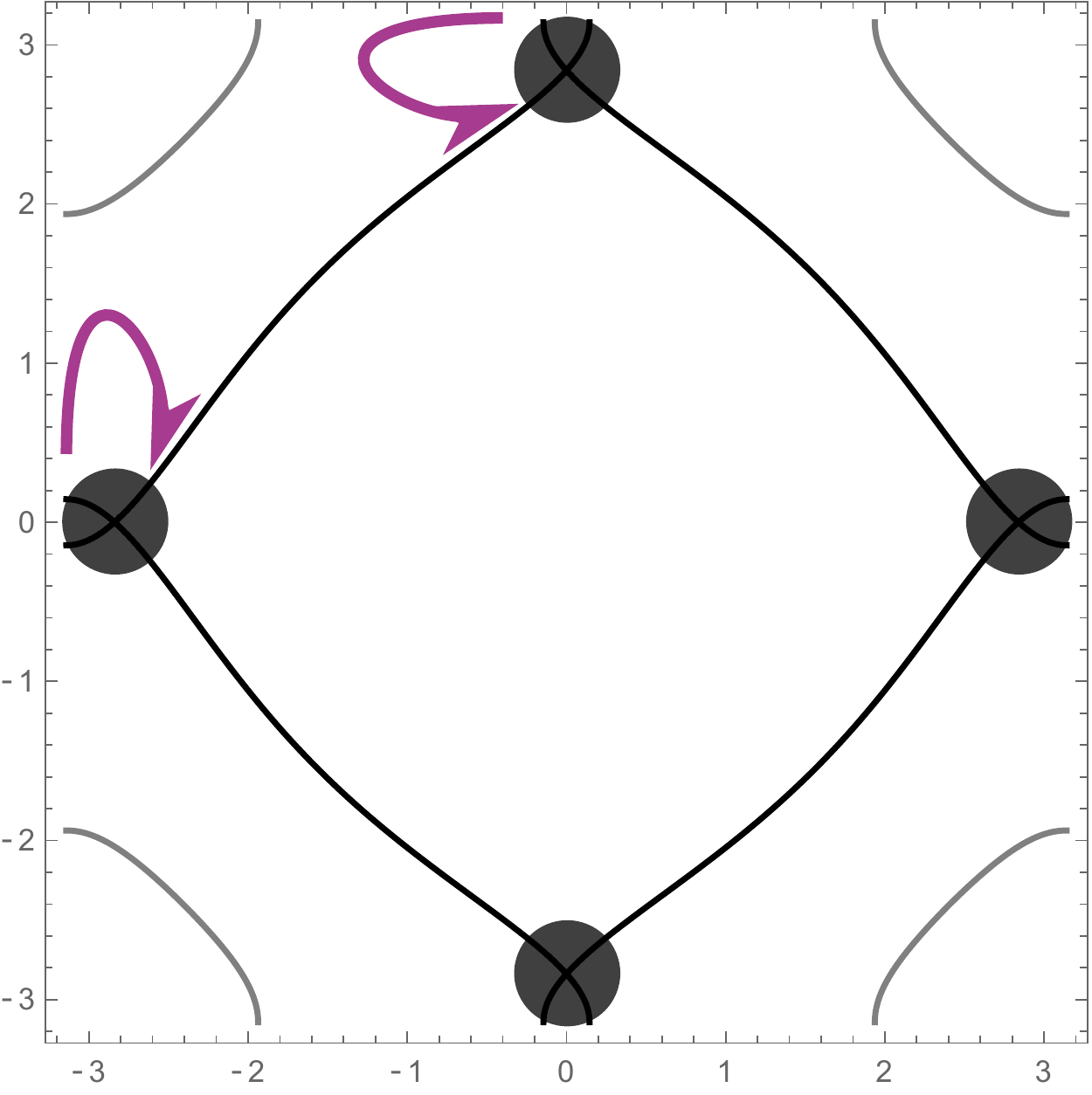}&	
	\includegraphics[width=0.32\columnwidth]{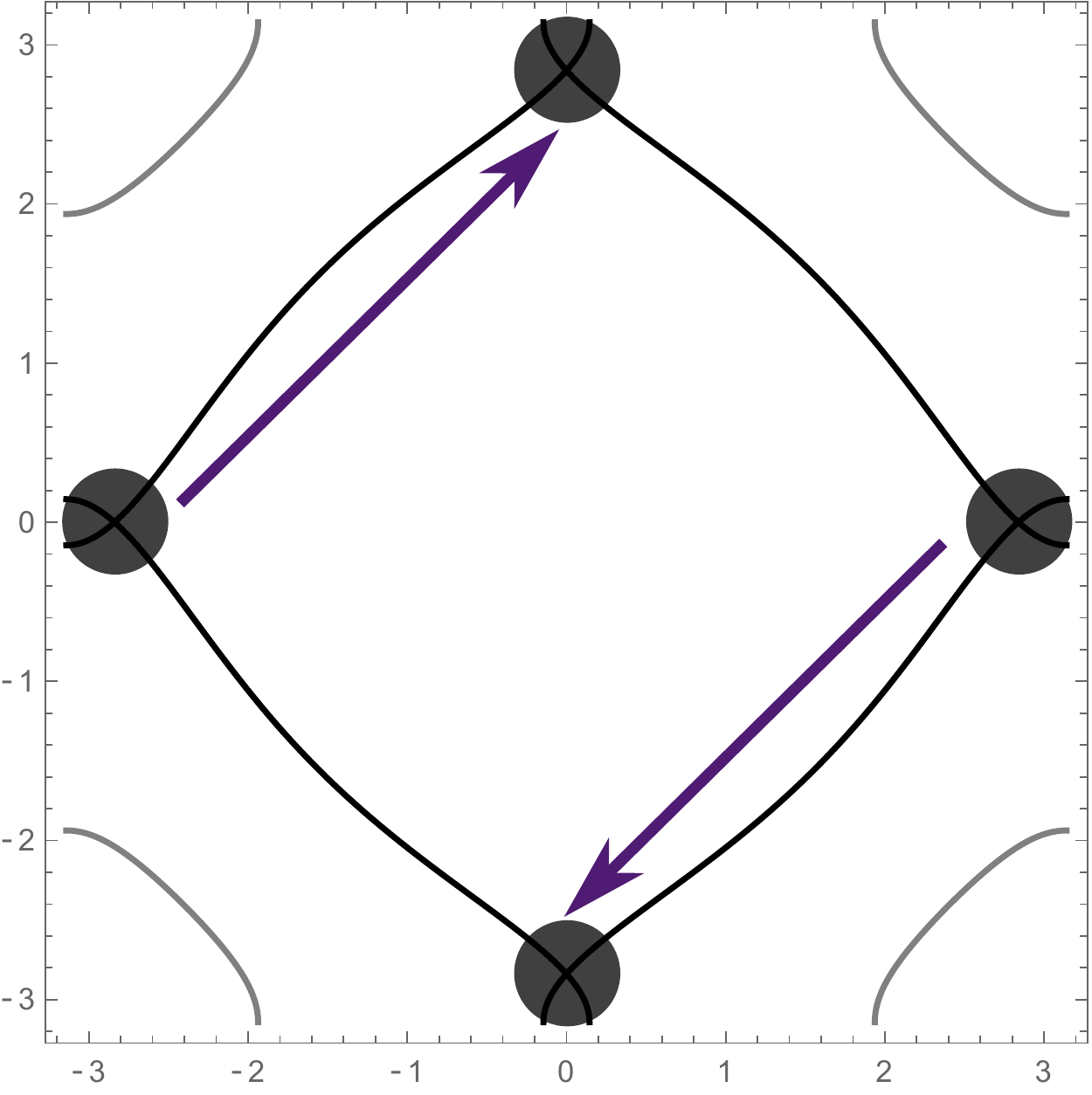}\\
	{$g_1$} & {$g_2$} & {$g_3$}
	\end{tabular}
	\caption{Interactions allowed with four patches and spinless fermions. Fermi surfaces at chemical potential $\mu_-$ top row and $\mu_+$ bottom row. $g_1$ and $g_2$ couplings denote density-density interactions between patches with zero and non-zero total momenta respectively. $g_3$ denotes an exchange interaction between all patches, conserving zero total momentum.}
	\label{ints}
\end{figure}

The low energy model that applies close to these van Hove fillings is given by the following imaginary time Lagrangian of spinless fermions,
\begin{align}
\mathcal{L}_\pm=&\sum_{a=1}^4\psi^{\dagger}_a(\partial_\tau-\xi^\pm_a(-i\partial_x,-i\partial_y))\psi_a-\frac{g_1}{2}\psi^{\dagger}_a\psi^{\dagger}_{\bar{a}} \psi_{\bar{a}}\psi_{a}\nonumber\\
&-\sum_{a=1}^{2}\sum_{b=3}^{4}g_2\psi^{\dagger}_a\psi^{\dagger}_{b}\psi_{b}\psi_a-\left[ig_3\psi^{\dagger}_4\psi^{\dagger}_{3} \psi_{2}\psi_1+\text{H.c.}\right].
\end{align}
We retain only states around the van Hove points within the theory due to their enhanced contribution to the low energy physics. This allows us to reduce the full Fermi surface to four regions with three possible interactions, depicted in Fig. \ref{ints}. We number the patches as shown in the left hand panels of Fig. \ref{ints}; $\bar{a}$ denotes the patch with opposite momentum to patch $a$. The couplings $g_i$ are marginal at tree level, which justifies the application of one-loop RG to the weakly coupled problem. Density-density interactions on the same patch have a momentum prefactor and are therefore irrelevant and discarded from the effective field theory. 

For the edge van Hove scenario there exist four inequivalent scattering vectors between patches. However, if we introduce four separate $g_2$ processes we find the couplings diverge equally and are indistinguishable during the flow. Therefore we describe the $g_2$ sector with one coupling for all scattering vectors.

The dispersions at the van Hove points are
\begin{eqnarray}\label{dispersions}
	& \displaystyle \xi^\pm_{1,2}(\mathbf{k})=-\frac{\delta k_x^2}{2m^\pm_x}+\frac{\delta k_y^2}{2m^\pm_y}, \quad \xi^\pm_{3,4}(\mathbf{k})=-\frac{\delta k_y^2}{2m^\pm_x}+\frac{\delta k_x^2}{2m^\pm_y}; & \nonumber \\
	& & \\
	& \displaystyle m_x^\pm=\pm\frac{1}{\sqrt{(t\pm2t')^2+\upsilon^2}}, & \\
	& \displaystyle m_y^\pm=\pm\frac{\sqrt{(t\pm2t')^2+\upsilon^2}}{t^2\pm2tt'+\upsilon^2+t\sqrt{(t\pm2t')^2+\upsilon^2}}. &
\end{eqnarray}
with $\delta k_x$ and $\delta k_y$ denoting the momentum relative to the van Hove saddle point value.

At van Hove filling, the density of states becomes logarithmic $\rho(\omega)\approx 2\lambda^{\pm}\ln(\Lambda /\omega)$ with $\Lambda $ an ultraviolet energy cutoff and $\omega$ the energy relative to the van Hove singularity. The constant $\lambda^{\pm}=\sqrt{m_x^\pm m_y^\pm}/4\pi^2$ for filling $\mu_\pm$. To determine the possible Fermi surface instabilities, the particle-particle and particle-hole susceptibilities are required. The susceptibilities that can have a double logarithmic divergence are\tcite{yao2015} 
\begin{align}\label{sus}
	&\chi_0^{\text{pp}}(\omega)\approx\lambda^{\pm}\ln^2\left(\frac{\Lambda }{\omega}\right), 
	&\chi_{\mathbf{q}_2}^{\text{ph}}(\omega)\approx2\beta^{\pm}\lambda^{\pm}\ln\left(\frac{\Lambda }{\omega}\right),
\end{align}
with
\beq
	\beta^{\pm}=\frac{2\sqrt{\kappa^{\pm}}}{1+\kappa^{\pm}}\ln\left|\frac{\kappa^{\pm}+1}{\kappa^{\pm}-1}\right|.
\eeq
 The ratio $\kappa^{\pm}=m^\pm_y/m^\pm_x$ plays the role of a nesting parameter with the logarithm in $\beta^{\pm}$  diverging as $\kappa\rightarrow1$ at perfect nesting. The complete set of relevant susceptibilities, including susceptibilities with single logarithmic divergences, are given in appendix \ref{fullsus}.

\section{Results}
We perform an RG analysis using $y=\ln^2(\Lambda /\omega)$ as a flow parameter with $\Lambda $ a decreasing energy cutoff\tcite{shankar1994,lehur2009}. The flow equations for the dimensionless couplings $g_i\rightarrow\lambda^{\pm} g_i$ are
\begin{align}
	&\dot{g}_1=-g_1^2-2g_3^2;\nonumber\\
	&\dot{g}_2=d(g_2^2+g_3^2);\nonumber\\
	&\dot{g}_3=-2g_1g_3+4dg_2g_3.
\end{align}	
The $y$-dependence of the $g_i$ has been suppressed for brevity. $\dot{g}_i$ denotes the derivative $dg_i/dy$. 
We have discarded contributions with single logarithmic divergences in this picture; the full flow equations are given in appendix \ref{fullrg}. $d\approx d\chi_{\mathbf{q}_2}^{\text{ph}}(y)/d\chi_0^{\text{pp}}(y)$ is approximated as a constant nesting parameter $0\leqslant d\leqslant1$ to account for the additional logarithmic divergence at perfect nesting. At the beginning of the flow $g_i(y=0)=\lambda^{\pm} U$ with all couplings equal.

\begin{figure}[t]
	\centering
	$\includegraphics[width=0.48\columnwidth]{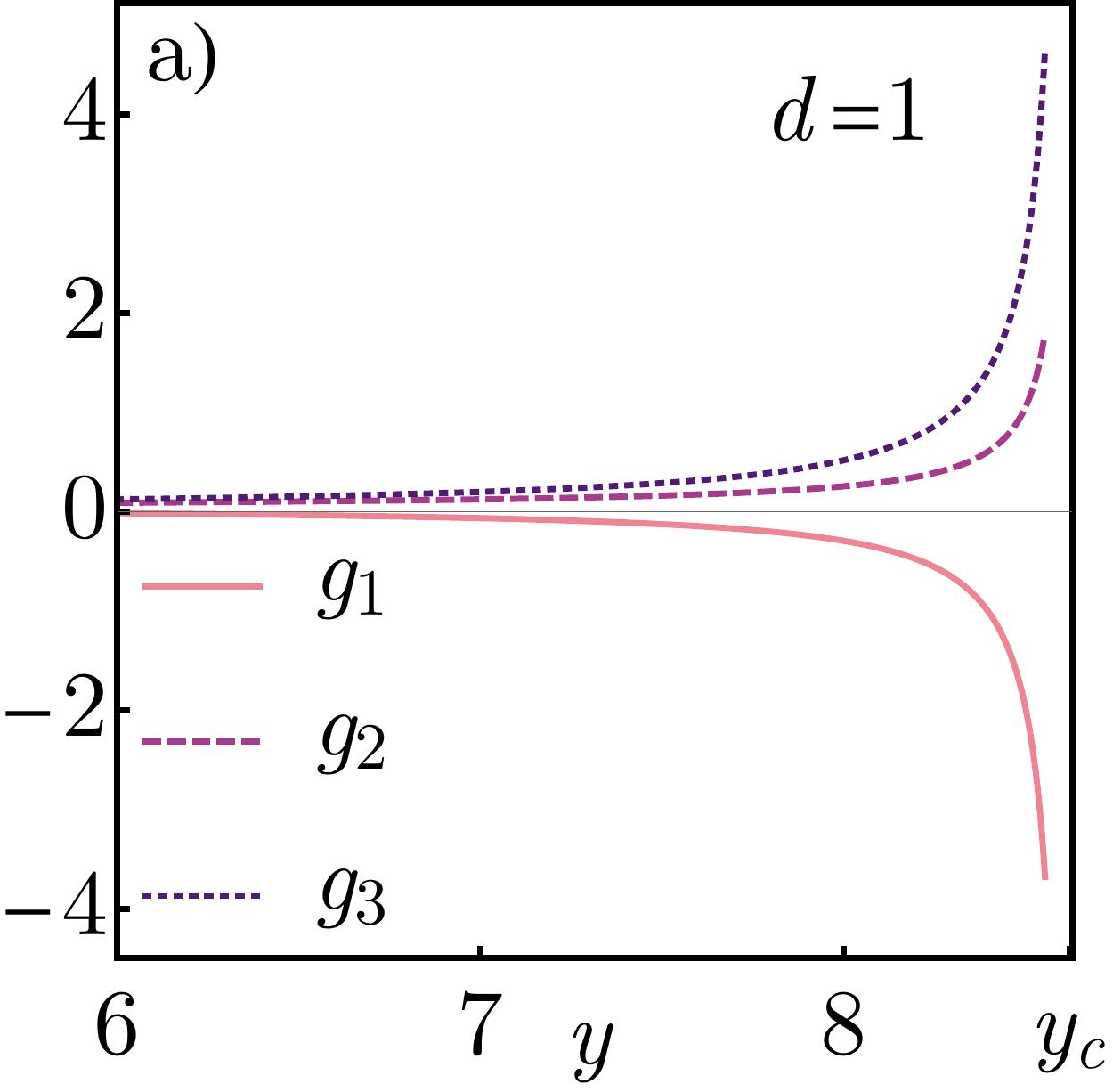}$\quad
	$\includegraphics[width=0.48\columnwidth]{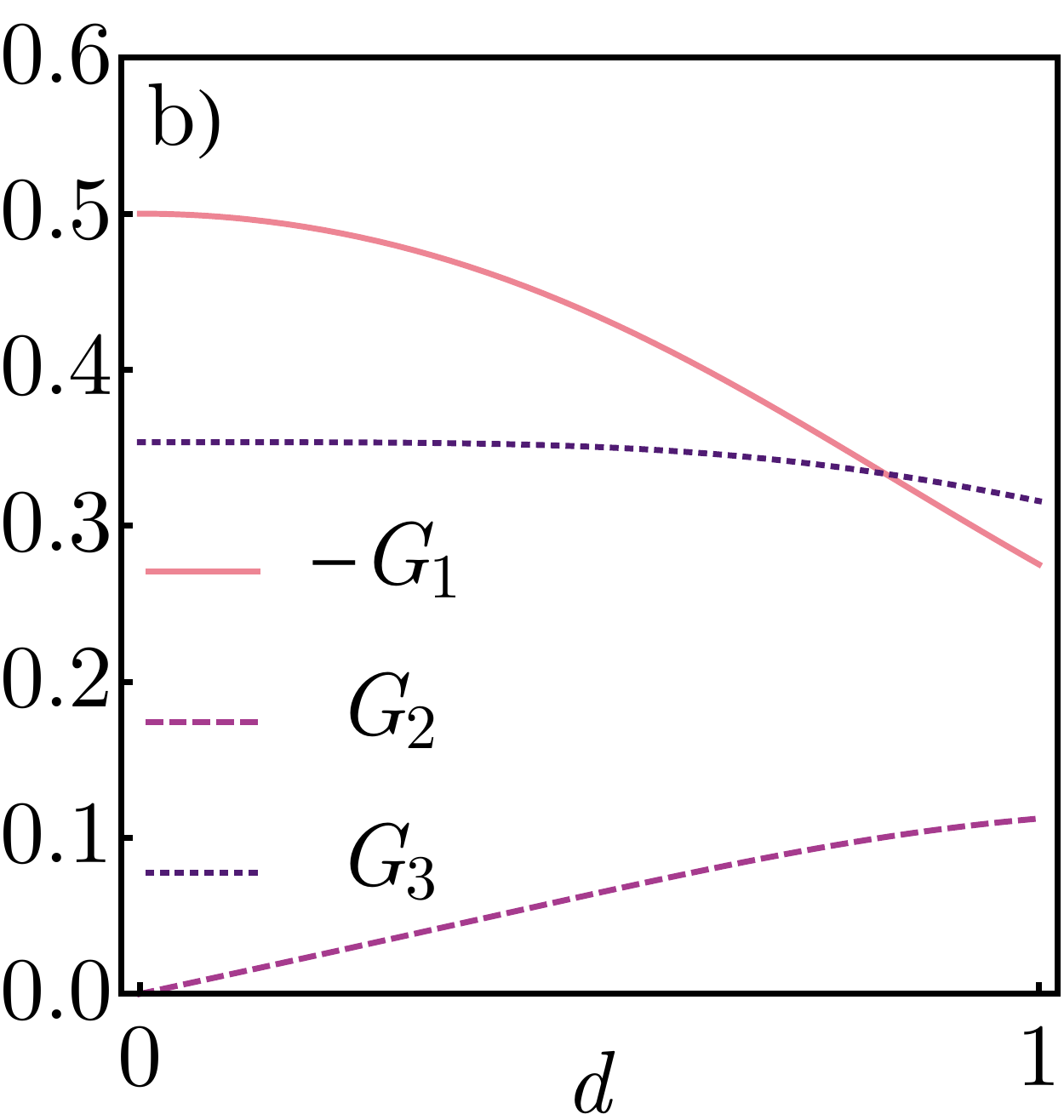}$
	\caption{a) Nesting parameter $d=1$ numerical solution for flow of couplings $g_i$ with RG scale $y$ at $\mu_+$ van Hove filling, starting from bare values $g_i(0)=0.04$. b) Exponent values in (\ref{gdiv}) for $0\leqslant d\leqslant1$.}
	\label{flowgi}
\end{figure}

For $d=0$ the differential equations can be solved analytically with the critical value $y_c=(1+\sqrt{2})/\lambda U$ for which the couplings diverge to strong coupling. We therefore use this as the cutoff for the phase transition. For $d\neq0$ the critical value decreases and the $g_3$ coupling is enhanced. The solutions to the RG equations for $d=0$ and $d=1$ are shown in Fig. \ref{flowgi}. $g_3$ retains the sign of $g_3(0)$ due to the $\beta$-function vanishing as the coupling goes to zero. $g_1$ decreases under the RG and eventually becomes negative, leading to superconductivity. 

The coupling constants $g_i$ flow to strong coupling as $y\rightarrow y_c$, therefore our one-loop RG can only provide a qualitative picture of the phase diagram. We introduce the asymptotic form
\beq\label{gdiv}
g_i\approx\frac{G_i}{y_c-y}
\eeq
to describe the divergence of the couplings\tcite{furukawa1998}.

To analyze the nature of the Fermi surface instabilities, we introduce infinitesimal test vertices for several possible types of order: superconductivity, $\mathbf{q}_1$ and $\mathbf{q}_2$ density waves, and Fulde-Ferrell-Larkin-Ovchinnikov (FFLO) superconductivity with finite momentum Cooper pairs. The resulting addition to the Lagrangian is
\begin{align}
\delta\mathcal{L}=&\sum_{a=1}^4[\Delta_{a\bar{a}}\psi^{\dagger}_a\psi^{\dagger}_{\bar{a}}+\phi_{a\bar{a}}\psi^{\dagger}_a\psi_{\bar{a}}]\nonumber\\&+\sum_{a=1}^2\sum_{b=3}^4[\phi_{ab}\psi^{\dagger}_a\psi_{b}+\Delta_{ab}\psi^{\dagger}_a\psi^{\dagger}_{b}]+\text{H.c.}
\end{align}
Spatially uniform ($\mathbf{q}=0$) charge and magnetic orderings are suppressed due to the irrelevance of the intra-patch density-density interaction.

We find for the superconducting channel
\beq 
\begin{pmatrix}
\dot{\Delta}_{12}	\\ 
\dot{\Delta}_{21}	\\ 
\dot{\Delta}_{34}	\\ 
\dot{\Delta}_{43}	
\end{pmatrix}
=\begin{pmatrix}
-g_1& g_1 & 2ig_3 & -2ig_3\\ 
g_1 & -g_1 & -2ig_3 & 2ig_3\\ 
-2ig_3 & 2ig_3 & -g_1 & g_1\\ 
2ig_3 & -2ig_3 & g_1 & -g_1
\end{pmatrix}
\begin{pmatrix}
{\Delta}_{12}	\\ 
{\Delta}_{21}	\\ 
{\Delta}_{34}	\\ 
{\Delta}_{43}	
\end{pmatrix}.
\eeq
 \begin{figure}[t]
	\centering
	\includegraphics[width=0.48\columnwidth]{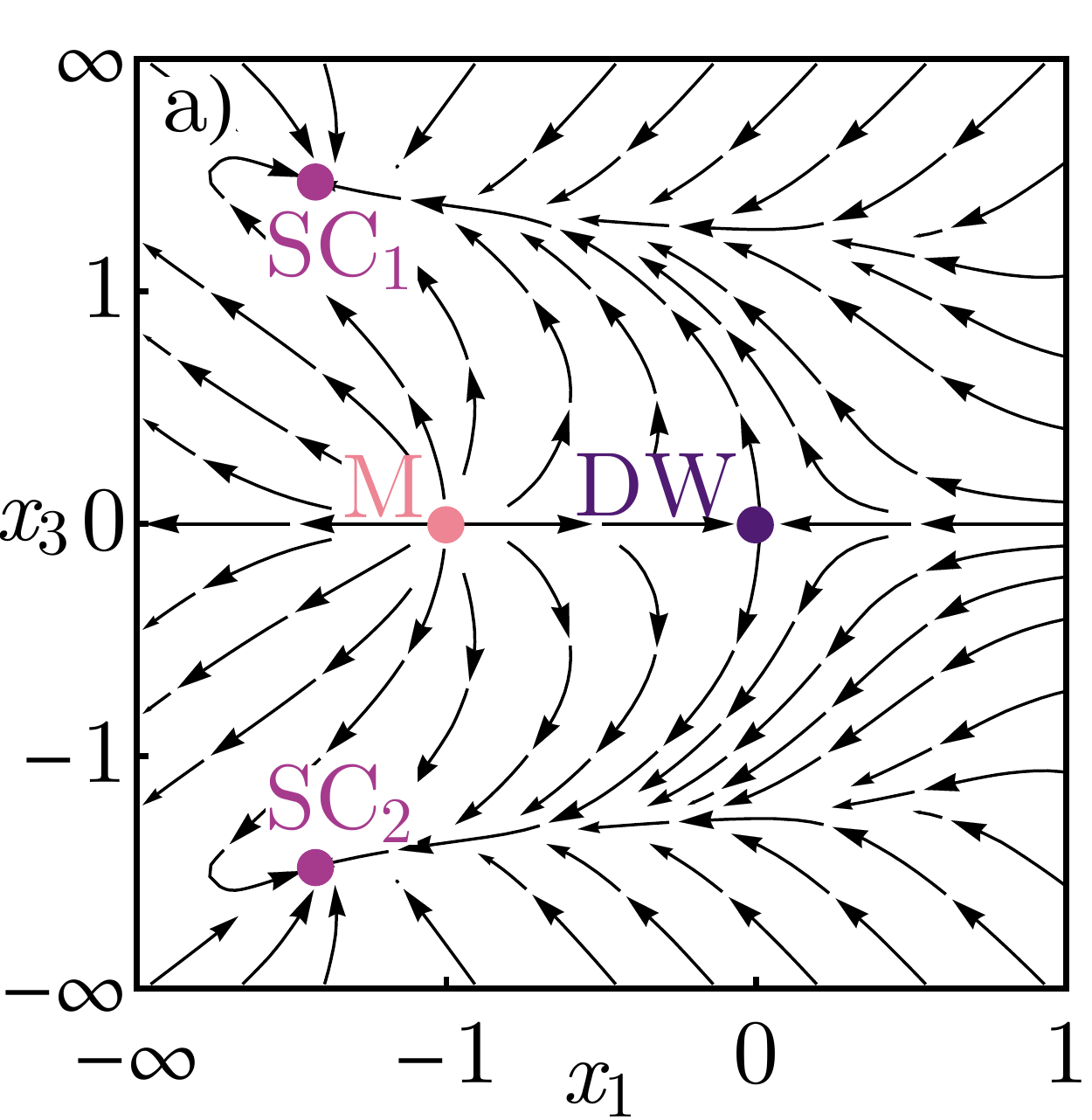}\;
	\includegraphics[width=0.48\columnwidth]{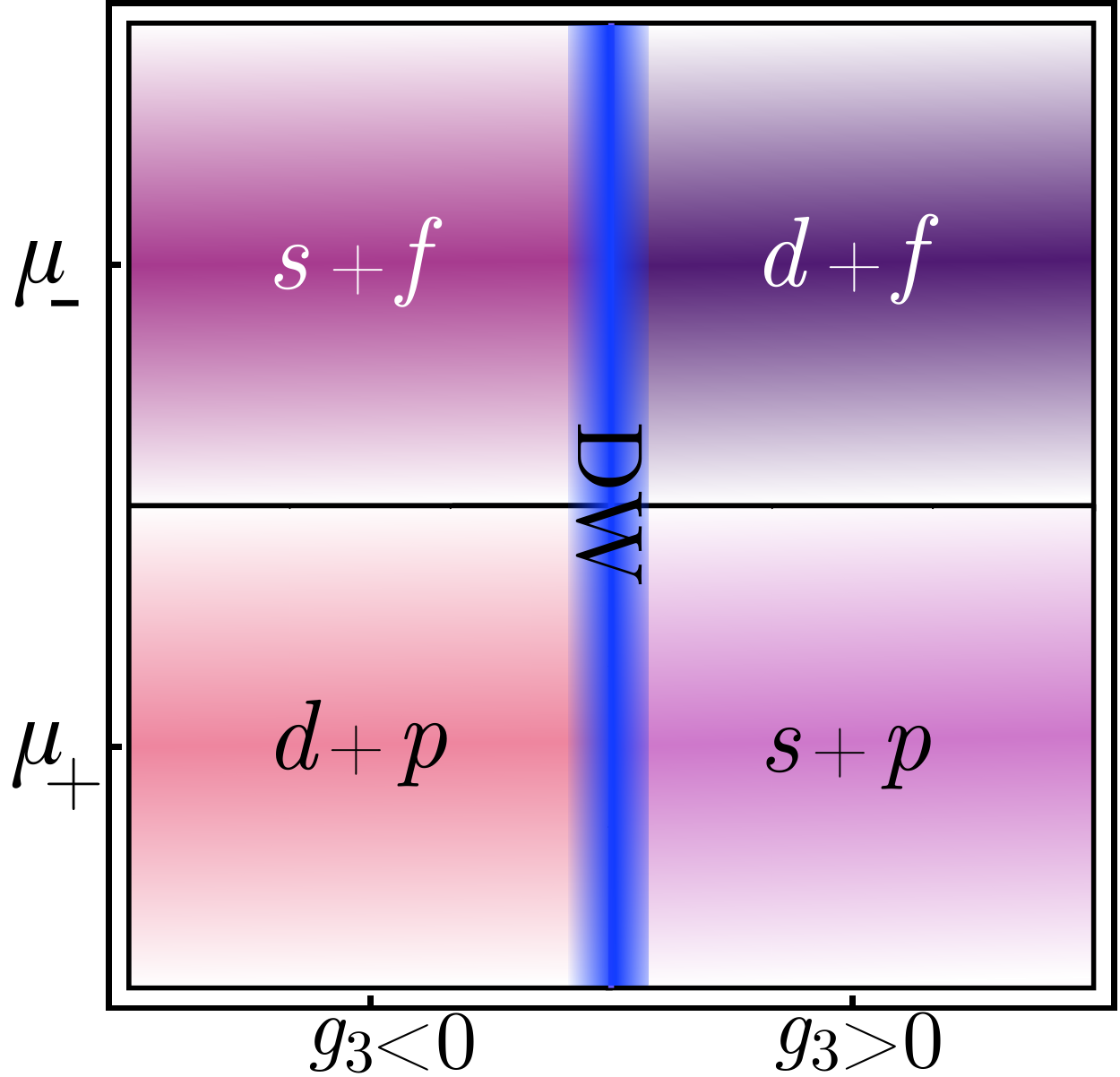}
	\caption{a) The RG flows of the couplings to log-squared accuracy and for nesting parameter $d=1$, projected onto the $x_1 x_3$-plane, where $x_1 = g_1/g_2$ and $x_3 = g_3/g_2$.  Fixed points in this diagram correspond to flow trajectories in the full space along which the ratios $x_1$ and $x_3$ become fixed.  To include the points at infinity, the axes have been rescaled according to $x \to x/(1+\vert x \vert)$.  M denotes the metallic trajectory, DW the $\textbf{q}_2$ density wave trajectory, and SC$_1$ and SC$_2$ the superconducting trajectories.  b) Schematic phase diagram near the two van Hove fillings, corresponding to chemical potentials $\mu_-$ and $\mu_+$ respectively.  For the model (\ref{rhham}), in which the on-site Hubbard interaction is repulsive, the coupling $g_3$ is always positive; however, for completeness we have included in the phase diagram also negative values of $g_3$, which might arise as a result of competing interactions in more complicated microscopic models.  The sign of $g_3$ determines whether the singlet component of the superconductivity is $s$-wave or $d$-wave.  Density-wave order dominates when $g_3$ is close to zero.}
	\label{phases}
\end{figure}

The dot again denotes a derivative with respect to $y$. The two possible nonzero eigenvalues of this matrix are $\varepsilon_{1}=-2(g_1-2g_3)$ and $\varepsilon_{2}=-2(g_1+2g_3)$ with corresponding eigenvectors $\mathbf{v}_1=(-i,i,-1,1)^\text{T}/\sqrt{4}$, $\mathbf{v}_2=(i,-i,-1,1)^\text{T}/\sqrt{4}$. The superconductivity in the helicity basis is chiral/anti-chiral depending on the sign of $g_3$. We repeat the analysis in the FFLO and the $\mathbf{q}_1$ and $\mathbf{q}_2$ density wave channels to find all possible orders. The order parameters obey $\dot{\Delta}_{j}=\varepsilon_j\Delta_j$; the susceptibilities of the possible orders are $\chi_j(y)\sim(y_c-y)^{G_j}$\tcite{furukawa1998}. 

The exponents for superconductivity and $\mathbf{q}_2$ density wave orders are given by $G_{\text{SC}_1}=2(G_1-2G_3)$, $G_{\text{SC}_2}=2(G_1+2G_3)$, $G_{\text{DW}\pm}^{\mathbf{q}_2}=-d(G_2\pm 2G_3)$. The FFLO superconductivity and $\mathbf{q}_1$ density wave order are suppressed and the $\mathbf{q}_2$ density wave order is also suppressed away from perfect nesting. The exponents for FFLO and $\mathbf{q}_1$ density wave order are $G_\text{FFLO}=2d_\alpha(y_c)G_2$, $G_\text{DW}^{\mathbf{q}_1}=-d_\gamma(y_c)G_1$ with $d_\alpha$ and $d_\gamma$ defined in appendix~\ref{fullrg}.

In order to obtain a picture of the RG flow to strong coupling we use the monotonically increasing $g_2$ as a flow parameter and redefine the remaining couplings $g_1=x_1 g_2$ and $g_3=x_3 g_2$\tcite{nandkishore2012}. The flow equations in terms of these redefined couplings are
\begin{align}
	&\frac{dx_1}{d\ln g_2}=-x_1-\frac{x_1^2+2x_3^2}{d(1+x_3^2)},\nonumber \\
	&\frac{dx_3}{d\ln g_2}=-x_3-\frac{2x_3(x_1-2d)}{d(1+x_3^2)}.
\end{align}
The fixed points of these equations describe four trajectories of the RG flow. The flow diagram is plotted in Fig. \ref{phases}a with $d=1$. For the metallic fixed point $g_2$ does not flow. The density wave phase exists for $g_3\rightarrow0$; in this case $g_1\rightarrow0$ and only $g_2$ diverges. For the superconducting trajectories all couplings diverge with ratios that depend on $d$. As $d\rightarrow0$, the density wave and metal trajectories merge and only the metallic phase survives.

We now consider the superconducting order parameter in the original $\{\uparrow,\downarrow\}$ spin basis. We name the discrete order parameter in analogy to the continuum angular momentum channels. The even-parity order parameters are the isotropic $s$-wave channel $\Delta_s=\Delta(1,1,1,1)$ and nodal $d$-wave  $\Delta_d=\Delta(1,1,-1,-1)$, where these four-component vectors give the phase of the superconducting order parameter at each of the four van Hove points, $(\mathbf{K}_1,\mathbf{K}_2,\mathbf{K}_3,\mathbf{K}_4)$.  The odd-parity order parameters correspond to chiral $p$-wave $\Delta_{p_x+ip_y}=\Delta(-i,i,-1,1)$, and anti-chiral $p$-wave $\Delta_{p_x-ip_y}=\Delta(i,-i,-1,1)$ in the Yao-Yang van Hove scenario but can represent higher order angular momentum channels for the edge van Hove scenario. The continuum order parameter can be written as $\bm{\Delta}(\mathbf{k})=(\Psi_s(\mathbf{k})+\mathbf{d}(\mathbf{k})\cdot\bm{\sigma})i\sigma_y$. In the Yao-Yang scenario the singlet component of the superconductivity $\Psi_s(\mathbf{k})$ corresponds to an $s$-wave form for $G_{\text{SC}_1}$ and a $d$-wave form for $G_{\text{SC}_2}$. The triplet component is helical and forms with chiral $p_x+ip_y$ superconductivity for one spin polarization and anti-chiral $p_x-ip_y$ superconductivity for the other, with $\mathbf{d}(\mathbf{k})=(\sin k_x,\sin k_y,0)^\text{T}$.

In the edge van Hove scenario the form factor is more complicated. If the van Hove points lie at $(\pm\pi/2,\pi)$, $(-\pi,\pm\pi/2)$ along the Brillouin zone edge, $G_{\text{SC}_1}$ corresponds to a superposition of singlet $d$-wave superconductivity with form factor $\Psi_s(\mathbf{k})=\cos 2k_x-\cos 2k_y$ and triplet $f$-wave superconductivity with the form factor $\mathbf{d}(\mathbf{k})=(\cos k_x-\cos k_y)(\sin k_x,\sin k_y,0)^\text{T}$. The singlet component of $G_{\text{SC}_2}$ is $s$-wave instead of $d$-wave. For van Hove points lying at different positions along the Brillouin zone edge the form factor requires higher harmonics, up to infinite order as the van Hove points approach the $(0,\pi)$ limit. 

Additionally there exists a narrow window of density wave order for $g_3\rightarrow0$. The competition between unconventional superconductivity and density-wave order has been seen to arise theoretically in similar spin-orbit split systems such as at oxide interfaces\tcite{scheurer2015}. A schematic phase diagram is given in Fig.~\ref{phases}b.

\section{Summary and discussion}
When the Fermi surface passes through saddle points in the band structure, the density of states is enhanced within the region of the saddle. This allows for an analytical treatment of the RG flow equations and an unbiased analysis of competing phases. 

We have shown that mixed-parity superconductivity arises naturally in systems with antisymmetric spin-orbit coupling. The direction of the triplet $\mathbf{d}(\mathbf{k})$ vector is determined by the local spin quantization axis. Thus the triplet component of the superconducting order parameter forms a helical state, analogous to the quantum spin Hall insulator\tcite{qi2009}. The helical superconductivity preserves time-reversal symmetry. The mixed-parity superconducting state is topologically non-trivial if the triplet component is greater than the singlet component\tcite{sato2009,lu2010}. Our case, where both components are equal, lies on the boundary between the topologically trivial and non-trivial phases. Our results suggest the superconductor can be tuned to a topological state, and could be useful for device applications and topological quantum computing. 

Recently we became aware of a related study on the hexagonal lattice\tcite{qin2019}.

\section{Acknowledgments} We thank J. Betouras and D. Efremov for useful discussions. MJT acknowledges financial support from the CM-CDT under EPSRC (UK) grant number EP/L015110/1. CAH acknowledges financial support from the EPSRC (UK), grant number EP/R031924/1.  He is also grateful to Rice University for a recent visiting position during which part of this work was completed.

\appendix

\section{Full particle-hole and particle-particle susceptibilities}\label{fullsus}
To compute the full RG flow equations including terms of single logarithmic divergence, all particle-particle and particle-hole susceptibilities are required. 

The complete expressions, including susceptibilities reproduced from (\ref{sus}), are\tcite{yao2015}:
\begin{align}
	&\chi_0^{\text{pp}}(\omega)\approx\lambda^{\pm}\ln^2\left(\frac{\Lambda }{\omega}\right), &\chi_{0}^{\text{ph}}(\omega)\approx2\lambda^{\pm}\ln\left(\frac{\Lambda }{\omega}\right),\nonumber\\
	&\chi_{\mathbf{q}_1}^{\text{ph}}(\omega\approx2\gamma\lambda^{\pm}\ln\left(\frac{\Lambda }{\omega}\right),
	&\chi_{\mathbf{q}_2}^{\text{ph}}(\omega)\approx2\beta^{\pm}\lambda^{\pm}\ln\left(\frac{\Lambda }{\omega}\right),\nonumber\\
	&\chi_{\mathbf{q}_2}^{\text{pp}}(\omega)\approx2\alpha^{\pm}\lambda^{\pm}\ln\left(\frac{\Lambda }{\omega}\right);
\end{align}
\begin{align}
	&\alpha^{\pm}=\frac{1+\kappa^{\pm}}{2\sqrt{\kappa^{\pm}}},&\beta^{\pm}=\frac{2\sqrt{\kappa^{\pm}}}{1+\kappa^{\pm}}\ln\left|\frac{\kappa^{\pm}+1}{\kappa^{\pm}-1}\right|.
\end{align}
The vectors $\mathbf{q}_1$ and $\mathbf{q}_2$ are given by $2\mathbf{K}_1$ and $\mathbf{K}_3-\mathbf{K}_1$ respectively. The $\pm$ signs of susceptibilities have been suppressed. $\gamma$ denotes an additional nesting parameter introduced by hand to suppress or enhance $\mathbf{q}_1$ scattering processes relative to the zero-momentum particle-hole processes\tcite{nandkishore2012}.

\section{Full RG flow equations}\label{fullrg}
The complete flow equations including all single and quadratic logarithmic terms are
\begin{align}\label{fullde}
	\dot{g}_1&=-g_1^2-2g_3^2-2g_2^2d_1+g_1^2d_\gamma;\nonumber\\
	\dot{g}_2&=-2g_1g_2d_1+(g_2^2+g_3^2)d_\beta-g_2^2d_\alpha;\nonumber\\
	\dot{g}_3&=-2g_1g_3+4g_2g_3d_\beta.
\end{align}	
The derivative $\dot{g}_i=dg_i/dy$. The $d_x(y)$ parameters are defined as $d_1(y)=d\chi_0^{\text{ph}}(y)/d\chi_0^{\text{pp}}(y)$, $d_\gamma(y)=d\chi_{\mathbf{q}_1}^{\text{ph}}(y)/d\chi_0^{\text{pp}}(y)$, $d_\beta(y)=d\chi_{\mathbf{q}_2}^{\text{ph}}(y)/d\chi_0^{\text{pp}}(y)$,  $d_\alpha(y)=d\chi_{\mathbf{q}_2}^{\text{pp}}(y)/d\chi_0^{\text{pp}}(y)$.

The functions $d_x(y)$, $x=1,\alpha,\beta,\gamma$; have the asymptotic forms $d_x(y)\rightarrow1$ as $y\rightarrow0$ and $d_x(y)\rightarrow x/\sqrt{y}$, for $y\rightarrow\infty$.

When solving the system of differential equations numerically we approximate the functions $d_x(y)$ by $d_x(y)=x/\sqrt{x^2+y}$ to interpolate between small-$y$ and large-$y$ asymptotic forms\tcite{nandkishore2012,yao2015}.

\end{document}